\documentstyle[12pt]{article}
\input epsf
\catcode`\@=11
\long\def\@makefntext#1{
\protect\noindent \hbox to 3.2pt {\hskip-.9pt
$^{{\ninerm\@thefnmark}}$\hfil}#1\hfill}		

\def\@makefnmark{\hbox to 0pt{$^{\@thefnmark}$\hss}}  

\def\ps@myheadings{\let\@mkboth\@gobbletwo
\def\@oddhead{\hbox{}
\rightmark\hfil\ninerm\thepage}
\def\@oddfoot{}\def\@evenhead{\ninerm\thepage\hfil
\leftmark\hbox{}}\def\@evenfoot{}
\def\sectionmark##1{}\def\subsectionmark##1{}}

\renewcommand{\thefootnote}{\fnsymbol{footnote}}

\newcounter{sectionc}\newcounter{subsectionc}\newcounter{subsubsectionc}
\renewcommand{\section}[1] {\vspace*{0.6cm}\addtocounter{sectionc}{1}
\setcounter{subsectionc}{0}\setcounter{subsubsectionc}{0}\noindent
	{\normalsize\bf\thesectionc. #1}\par\vspace*{0.4cm}}
\renewcommand{\subsection}[1] {\vspace*{0.6cm}\addtocounter{subsectionc}{1}
	\setcounter{subsubsectionc}{0}\noindent
	{\normalsize\it\thesectionc.\thesubsectionc. #1}\par\vspace*{0.4cm}}
\renewcommand{\subsubsection}[1]
{\vspace*{0.6cm}\addtocounter{subsubsectionc}{1}
	\noindent {\normalsize\rm\thesectionc.\thesubsectionc.\thesubsubsectionc.
	#1}\par\vspace*{0.4cm}}

\newcounter{appendixc}
\newcounter{subappendixc}[appendixc]
\newcounter{subsubappendixc}[subappendixc]

\renewcommand{\appendix}[1] {\vspace*{0.6cm}
        \refstepcounter{appendixc}
        \setcounter{figure}{0}
        \setcounter{table}{0}
        \setcounter{equation}{0}
        \renewcommand{\thefigure}{\Alph{appendixc}.\arabic{figure}}
        \renewcommand{\thetable}{\Alph{appendixc}.\arabic{table}}
        \renewcommand{\theappendixc}{\Alph{appendixc}}
        \renewcommand{\theequation}{\Alph{appendixc}.\arabic{equation}}
        \noindent{\bf Appendix \theappendixc #1}\par\vspace*{0.4cm}}

\def\abstracts#1{{

\centering{\begin{minipage}{12.2truecm}\footnotesize\baselineskip=12pt\noindent
	\centerline{\footnotesize ABSTRACT}\vspace*{0.3cm}
	\parindent=0pt #1
	\end{minipage}}\par}}

\renewenvironment{thebibliography}[1]
	{\begin{list}{\arabic{enumi}.}
	{\usecounter{enumi}\setlength{\parsep}{0pt}
\setlength{\leftmargin 1.25cm}{\rightmargin 0pt}
	 \setlength{\itemsep}{0pt} \settowidth
	{\labelwidth}{#1.}\sloppy}}{\end{list}}

\newcounter{itemlistc}
\newcounter{romanlistc}
\newcounter{alphlistc}
\newcounter{arabiclistc}

\newcommand{\fcaption}[1]{
        \refstepcounter{figure}
        \setbox\@tempboxa = \hbox{\footnotesize Fig.~\thefigure. #1}
        \ifdim \wd\@tempboxa > 6in
           {\begin{center}
        \parbox{6in}{\footnotesize\baselineskip=12pt Fig.~\thefigure. #1}
            \end{center}}
        \else
             {\begin{center}
             {\footnotesize Fig.~\thefigure. #1}
              \end{center}}
        \fi}

\newcommand{\tcaption}[1]{
        \refstepcounter{table}
        \setbox\@tempboxa = \hbox{\footnotesize Table~\thetable. #1}
        \ifdim \wd\@tempboxa > 6in
           {\begin{center}
        \parbox{6in}{\footnotesize\baselineskip=12pt Table~\thetable. #1}
            \end{center}}
        \else
             {\begin{center}
             {\footnotesize Table~\thetable. #1}
              \end{center}}
        \fi}

\def\@citex[#1]#2{\if@filesw\immediate\write\@auxout
	{\string\citation{#2}}\fi
\def\@citea{}\@cite{\@for\@citeb:=#2\do
	{\@citea\def\@citea{,}\@ifundefined
	{b@\@citeb}{{\bf ?}\@warning
	{Citation `\@citeb' on page \thepage \space undefined}}
	{\csname b@\@citeb\endcsname}}}{#1}}

\newif\if@cghi
\def\cite{\@cghitrue\@ifnextchar [{\@tempswatrue
	\@citex}{\@tempswafalse\@citex[]}}
\def\citelow{\@cghifalse\@ifnextchar [{\@tempswatrue
	\@citex}{\@tempswafalse\@citex[]}}
\def\@cite#1#2{{$\null^{#1}$\if@tempswa\typeout
	{IJCGA warning: optional citation argument
	ignored: `#2'} \fi}}

 1
 1
 1

\font\ninerm=cmr9

\begin{document}
\rightline{NSF-ITP-95-63}
\rightline{hep-th/9507094}
\vspace*{1.5cm}
\centerline{\normalsize\bf STRING THEORY}
\baselineskip=16pt
\centerline{\normalsize\bf AND BLACK HOLE COMPLEMENTARITY}

\vspace*{0.6cm}
\centerline{\footnotesize JOSEPH POLCHINSKI}
\baselineskip=13pt
\centerline{\footnotesize\it Institute for Theoretical Physics,
University of California}
\baselineskip=12pt
\centerline{\footnotesize\it Santa Barbara, CA 93106-4030}
\centerline{\footnotesize E-mail: joep@itp.ucsb.edu}

\vspace*{0.9cm}
\abstracts{Is string theory relevant to the black hole
information problem?  This is an attempt to clarify some of the issues
involved.}

\vspace*{0.6cm}
\normalsize\baselineskip=15pt
\setcounter{footnote}{0}
\renewcommand{\thefootnote}{\alph{footnote}}
In spite of the great effort that the black hole information problem has
inspired, the situation has in some ways changed little since the
original work of Hawking.$^1$  The three principal alternatives (that
information is lost, stored in a remnant, or emitted with the Hawking
radiation) remain, none having been convincingly ruled out or shown to be
consistent.  The issues have been sharpened, but there
is no consensus.

Perhaps the most novel proposal is the principle of
black hole complementarity$^2$ as realized in string theory.$^3$
This talk is about an attempt to understand these ideas, and is based on
work in collaboration with Lowe, Susskind, Thorlacius, and Uglum.$^4$

\section{The Nice Slice Argument}

Before we turn to string theory itself let us ask,
do we expect the Hawking radiation to depend on short distance (Planck
scale) physics?  Even on this basic question there are
vociferous differences of opinion, and it is easy to see why.  On the one
hand, the horizon of a macroscopic black hole is a very smooth place.  The
tidal forces need be no larger than in this room, and we would expect
therefore to be able to use low energy effective field theory.  On the
other hand, many (some would claim all) derivations of the Hawking
radiation make explicit reference to ridiculously large
energies, greater than the mass of the universe, and assume for example
that free field theory is valid at these energies.

So, can we derive the Hawking radiation in a way that takes advantage of
the smoothness of the geometry?  It seems that the right way to do this
is in a Hamiltonian framework, pushing forward the state of the system on
a series of spacelike surfaces.  In order to use low energy field theory
everywhere, the slices need to be smooth, without large curvatures or
accelerations, and any matter (the asymptotic observer, the
infalling body) must be moving with modest velocity in the local frame
defined by the slice.  We will refer to these as ``nice slices.''

To construct one family of nice slices, let us describe the
Schwarzschild
black hole in Kruskal-Szekeres coordinates, which we will call $(x^+,
x^-)$ rather than the usual $(U, V)$.  In these coordinates the
singularity is at $x^+ x^- = 16G^2 M^2$, and the event horizon is the
surface $x^+=0$ ($x^+$ increases to the upper left).  We construct a
spacelike surface composed of two pieces.  The first piece is the left
half ($x^- < x^+$) of the hyperbola $x^+ x^-  = R^2$.  This is chosen to
be inside the horizon but far from the singularity, so the geometry is
still smooth; for example let $R = 4G^2 M^2$.  The
second piece of the nice slice is the half-line $x^+ + x^- =
2R$ for $x^- > x^+$.  The slice is shown in fig.~1.  At large distance
this  slice
is asymptotic to the constant time surface $t = 0$.
The slice can be pushed forward and backward
in time by using the Killing symmetry of the black hole geometry,
\begin{eqnarray}
x^+ &\rightarrow&  x^+ e^{-t/4GM} \nonumber\\
x^- &\rightarrow&  x^- e^{t/4GM} .
\end{eqnarray}
Since the nice slices are asymptotic to surfaces of constant
Schwarzschild time, they can be parametrized by $t$.  The full set of
slices can then be written
\begin{eqnarray}
&&x^+ x^- = R^2 \>, \qquad x^- < e^{t/2GM} x^+ \>, \nonumber\\
&&e^{t/4GM} x^+ + e^{-t/4GM} x^- = 2R \>, \qquad x^- > e^{t/2GM} x^+ \>.
\label{nice}
\end{eqnarray}
The join between the line segment and the
hyperbola on each slice should be smoothed to avoid a large extrinsic
curvature gradient there.
\begin{figure}
\begin{center}
\leavevmode
\epsfbox{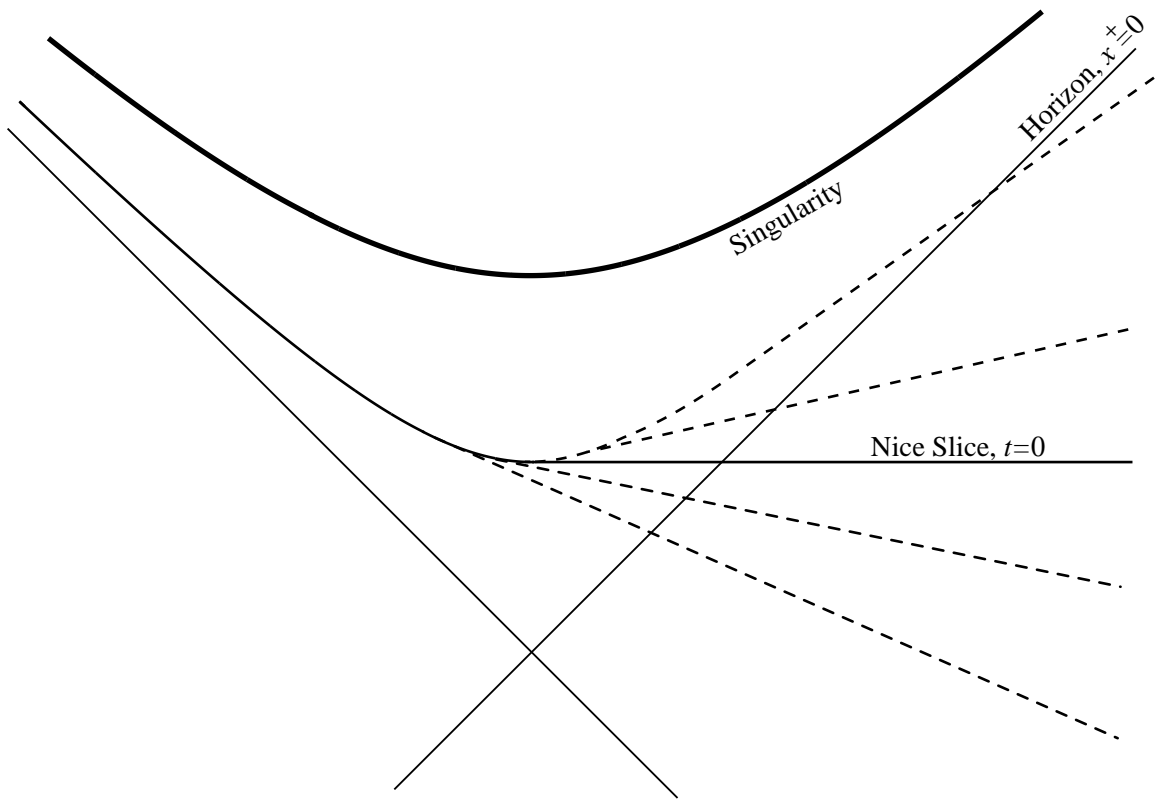}
\end{center}
\caption[]{Nice slicing of the Kruskal $x^+ x^-$ plane.  The slice which is
asymptotic to the $t=0$ surface is drawn as a solid line, with earlier
and later slices shown dashed.}
\end{figure}

It is not hard to check that the velocity (and energy) of an
infalling particle, as measured locally in terms of the time coordinate
orthogonal to the nice slice, remain small even as it passes through the
horizon.  For a black hole formed by collapse, one can join the nice
slices smoothly onto a set of smooth slices in the interior of the
collapsing body.  Also, as the black hole evaporates the background
geometry changes.  The nice slices can be adjusted along with the change
in the geometry  until very late in the evaporation when the curvature
becomes large.

Starting with the initial diffuse matter from which the black hole formed,
one can evolve the state of the system forward on such nice slices until
the evaporation is nearly complete and the curvature becomes large.  For a
large black hole, almost all of the original mass will have been
converted to Hawking radiation, which will be outgoing on the exterior
part of the nice slice.  By construction the geometry changes
smoothly from slice to slice, so the adiabatic theorem implies that only
very low-energy degrees of freedom ($E\sim 1/GM$) are excited
from their ground state in the Hawking emission process.
Thus, the state on the last nice slice is obtained from the initial
diffuse state using only low energy field theory.

It is of course rather clumsy to actually carry out the calculation in
the above fashion, but it is not necessary.  The essential features of
the final state can be deduced indirectly in two different ways.  First,
having argued that the result is independent of short distance physics,
we can make any convenient assumption about the form of the short
distance theory, in particular that it is simply free field theory.  The
result must then be the same as in the standard calculation: the
outgoing radiation looks thermal and is in fact in a highly mixed quantum
state, being correlated with fluctuations behind the horizon.\footnote{I
believe that this justification is implicitly assumed in the usual
derivation, but I do not know of anywhere in the literature that the
nice slice argument is given in detail.  Wald has pointed out the
existence of nice slices (private communication) and given some discussion
in ref.~5.}

Second, we can see this result directly as follows.  Consider the fields on
a distance scale of order one fermi, chosen to be very short
compared to the scale of the geometry but still in a region which physics
is well understood.  In particular, focus on the state of
these fields near the event horizon, at a distance again of order one
fermi.  `Distance' here is measured in the spacelike
direction defined by the nice slice.  The geometry is of course smooth at
the event horizon, and so the adiabatic theorem tells us that these modes
are unexcited.  The subsequent evolution in the curved background,
involving only length scales greater than a fermi, relates
the annihilation operators of the outgoing Hawking radiation to a linear
combination of the creation and annihilation operators for the fermi-scale
fields near the horizon.  The relation is such that the latter fields
being in their ground state implies that the asymptotic modes have a
thermal spectrum.

This establishes the gross thermal character of the Hawking radiation,
but we can extend the argument to learn more.  Consider two fermi-scale
wavepackets, one inside and one outside the horizon as shown
schematically  in fig.~2a.
\begin{figure}
\begin{center}
\leavevmode
\epsfbox{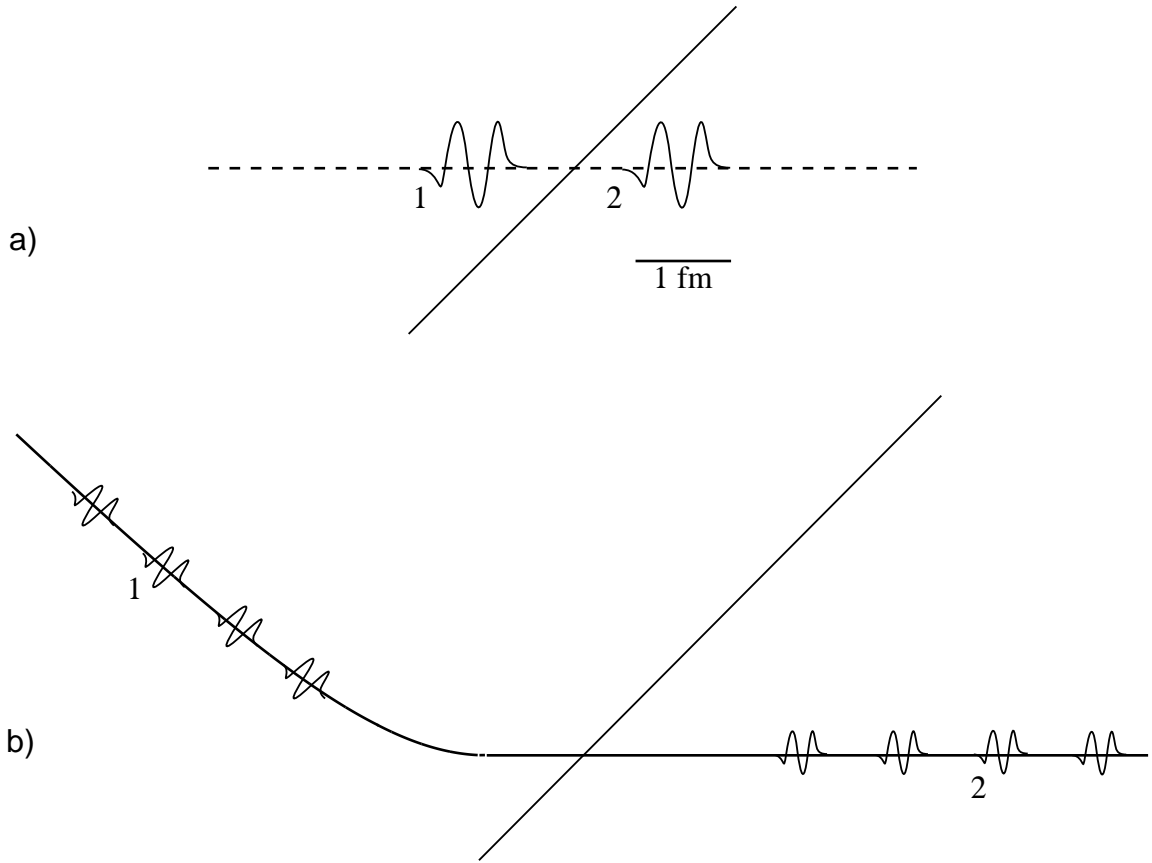}
\end{center}
\caption[]{a) Fermi-scale wavepackets just inside and just outside the
horizon.  b) Subsequent state of several pairs of packets along a nice
slice.}
\end{figure}
Let $\phi_1$ and $\phi_2$ denote some quantum
field averaged over these packets.  By the nature of the vacuum these are
correlated,
\begin{equation}
\langle \phi_1 \phi_2 \rangle - \langle \phi_1 \rangle
\langle \phi_2 \rangle \neq 0.   \label{corr}
\end{equation}
The geometry near the horizon is something like an expanding universe, so
that as time goes on the two packets redshift and separate while the
correlation remains.
Eventually their sizes, and the distance between them, are of order the
scale $M^{-1}$ of the geometry.  Fig.~2b shows the packets at this point,
together with some earlier and later pairs.
The nonvanishing correlation~(\ref{corr}) means that the outgoing
radiation is not in a pure state.  One can follow the evolution in this
way until the evaporation is nearly complete and the black hole not much
larger than Planck-sized.  At this point the state of the external
radiation is highly correlated with that of the quantum fields inside the
horizon and so is very far from a pure state.  The black
hole is now a small object which must have an enormous number of
internal states.

The later evolution depends on short distance physics.
The black hole might decay completely, leaving a mixed state.  It might
remain as an eternal remnant.  It might decay {\it extremely} slowly
leaving in the end a pure state with correlations between the Hawking
radiation and the final decay products.  The one thing that cannot happen
is that the purity of the final state is suddenly restored in the final
instants of Hawking evaporation: by a large margin there are too few
quanta to carry the necessary correlations (see ref.~6 for a
quantitative analysis).

One natural outcome is the interior of the black hole pinching off,
a change of topology.  In this case the final state from the point of
view of the exterior universe remains pure.  The decay Hamiltonian
involves a large finite number of new parameters, which are eigenvalues
of the third quantized baby universe field, and the topology changing
decay must again be extremely slow.$^7$  It has been argued that even
without reference to topology change, if the decay proceeds to completion
it must do so in essentially this way as seen from outside.$^8$

This analysis seems to leave no room for stringy short distance
behavior or any other modification of field theory to influence the
details of the Hawking radiation.
Yet the conclusion, that the Hawking radiation is in a highly mixed state
correlated with an almost Planck-sized black hole, is not entirely
appealing.

The next section addresses the question of whether the nice slice
analysis might fail in string theory, but let us first expand on
the preceding discussion.\footnote
{This is my side of discussions with Erik and
Herman Verlinde.}
Gravitational back reaction is a
nonrenormalizable interaction, and so should be irrelevant in the low
energy theory (except for the slow semiclassical evolution of the
background geometry).  However, there are claims that certain calculations
reveal a large effect.
To see what the issue is, consider quantum fields in a box which
which is slowly expanding.  The expansion continues until the box has
expanded by an enormous factor; in the black hole the expansion
factor will be $e^{O(M^2/M_{\rm P}^2)}$, with $M^2$ the original black
hole mass.  Consider also the reverse process, slow contraction by a large
factor.  Although both processes sound adiabatic, in only one of the two
cases can low energy field theory be used---the slow expansion.  Although
the change is slow, for massless modes of sufficiently low frequency the
adiabatic approximation breaks down and real quanta are produced.  In the
expanding box these simply redshift away, but in the contracting box they
will be blue-shifted until they reach a freqency where low energy field
theory breaks down.  So low energy field theory is valid in this geometry
only for time evolution in one direction.

The black hole is like the expanding box.  As we have noted in discussing
fig.~2, modes near the horizon redshift.  The time constant is $4GM$ and
the black hole lifetime is of order $GM^3/M_{\rm P}^2$, so the number of
potential $e$-foldings is large.  Fortunately, low energy field theory is
valid for the purpose we have applied it to in the nice-slice analysis,
obtaining the final state for a given initial state.  It is {\it not}
valid if we ask: given a black hole in a particular state on some late
nice slice, what was the initial state from which it evolved?  This
requires that we evolve backwards, so like the contracting box the answer
is outside the range of low energy field theory and likely involves some
very complicated state with large deviations from the semiclassical black
hole geometry.  As far as I can see, all claims of large back-reaction
effects involve such backwards questions, which are not relevant to the
nice-slice argument.

The low energy field theory has one unusual
feature.  Let us take the cutoff length at some scale $\ell$ which is
small compared to the scale of the geometry but large compared to the
Planck length, for example one fermi as we used in discussing fig.~2a.
As the box grows, the number of states below the cutoff, the number of
states for which low energy field theory is applicable, also grows.
So time evolution in the low energy effective theory must be from a smaller
Hilbert space into a larger.  This presents no problem.  As new degrees
of freedom enter the low energy theory by red-shifting through the scale
$\ell$, they are in their ground states due to the adiabatic theorem.
The evolution is thus well-defined and is one-way unitary (if the system
starts out in the low energy Hilbert space it ends up in the low energy
Hilbert space with probability essentially one, but the opposite is not
true).
Of course in the exact theory the evolution is assumed to be given by
ordinary quantum mechanics.  It is tempting to look at the large final
Hilbert space of low energy states and ask where those states `came from,'
but this is a backwards question and outside the range of validity of low
energy field theory.  Fortunately there is no need to answer it.

\section{Black Hole Complementarity}

We have argued that low energy field theory is valid and leads to a
certain conclusion.  This seems to leave little room for string theory.
To be precise, however, we have assumed that low energy field theory
knows its own range of validity.  In the black hole geometry there are no
large local invariants but there is a large non-local invariant, the
relative Lorentz boost of the infalling body and asymptotic observer.
At times of order the black hole lifetime~$GM^3/M_{\rm P}^2$,
the hyperbolic portion of the nice slice~(\ref{nice}) between the
infalling body and the asymptotic observer is very long.  Comparing frames
by parallel transport along the slice, one finds a rapidity difference of
order~$M^2/M_{\rm P}^2$. In field theory this large non-local invariant
does not cause any breakdown (for the forward evolution).  But it is a
logical possibility that when it is large low energy field theory ceases
to be a good approximation to string theory and we cannot use it.

In order to resolve the information problem, it is necessary that the
correlations between the interior and exterior fields on the late slice of
fig.~2b are somehow transmuted into correlations among the external
fields.  It is difficult to imagine a mechanism that would erase the
correlations between the internal and external fields, and the
superposition principle forbids the correlations from being duplicated in
an independent set of degrees of freedom.$^3$  The principle of black
hole complementarity works in a more subtle way.$^{2,3}$  That is, the
Hilbert space structure on the nice slice is supposed to be very different
from low energy field theory, so that the interior and outgoing fields are
actually {\it the same} degrees of freedom seen in very different
Lorentz frames.

Strings do have at least one unusual property at large boost,
transverse growth.$^{9}$  At large dilation factor $\gamma$ the
transverse size grows as $\sqrt{\ln \gamma}$, an effect which can be seen
both in the light-cone wavefunction and in the S-matrix.  A root-log
is very slow, but strikingly it combines with the exponential boost
factor $\gamma = e^{t/4GM}$ in the black hole geometry to give a diffusive
growth,$^{10}$ which is expected to be accelerated still further by
interactions.$^{11}$
This is not enough, however.  A non-local effect is needed {\it along} the
nice slice, in the longitudinal direction.  The light-cone wavefunction
does also show longitudinal spreading,$^{9}$ but it is much more
difficult to see this in the S-matrix or say whether it is enough to
resolve the information problem.

We need that the low energy degrees of freedom of the infalling observer
be secretly the same as those of the external observer.  The low energy
field operators of these two observers will then no longer commute, even
though they are at large spatial separation.  So let us calculate the
commutator.  Consider two spacetime points, $x_1$ and
$x_2$, which lie on a fixed nice slice corresponding
Schwarzschild time $t$.  The time $t$ is chosen large, but not so
large that an appreciable amount of evaporation has occurred.  The
point $x_2$ lies behind the horizon, and could be chosen to lie on
the hyperbola $x^+ x^-=R^2$.  It may be thought of as a point near the
trajectory of a low-energy particle which has fallen through the
horizon at some early time.  Point $x_1$ lies outside the event
horizon at
\begin{equation}
x^+_1 = x^+_0 e^{-t/4GM}, \qquad
x^-_1 = x^-_0 e^{t/4GM},
\end{equation}
with $x^+_0 < 0$, $x^-_0 > 0$, $x^+_0 + x^-_0 = 2R$.
For $x^+_0 x^-_0 = \alpha'$
this point is on the `stretched horizon,' where
the infalling information is supposed to be stored, according to the
reckoning of an observer who stays outside the black hole.

As $t$ increases, the spacelike separation between $x_1$ and $x_2$
grows like $e^\omega = \exp (t/4GM)$.  A field which has momenta of
order~1 in the nice slice frame at $x_2$ has momentum components
$p^+ = O(e^{-\omega})$, $p^- = O(e^{\omega})$.  So we wish to evaluate
the commutator for
the mass eigenstate component fields $\phi(x)$
of the string and then fold into suitable wavepackets.
We need a Hilbert space description of string theory
and so will use light-cone string field theory.\footnote
{The commutator has previously been studied in refs.~12, and in
particular the transverse spreading was seen in ref.~13.}
  This has not been
extended to the black-hole geometry, but for a large black hole it should
be sufficient to consider the flat space-time commutator, and ask whether
local commutativity breaks down at large boost.

We actually calculate the square of the commutator,
\begin{equation}
\langle 0| [{ \phi}(x_1),{ \phi}(x_2)][{ \phi}({x_2}'),{ \phi}({x_1}')]|0
\rangle\ .
\end{equation}
This is essentially local in free string field theory$^{12}$ and gets its
first interesting contribution from second order perturbation theory.
The details are left to ref.~4.  The result is that the commutator is
indeed nonlocal.  In fact it {\it grows} as $e^{\omega(\alpha(t) - 1)}$
where $\alpha(t) = 2 + \alpha' t/4$ is the closed string Regge trajectory
(here $t$ is the Mandelstam variable).  Moreover the typical
intermediate state contributing is a long string stretching between $x_1$
and $x_2$.  This is not in the naive low energy field theory, but appears
in the commutator of two low energy fields on the nice slice.

This is just as black hole complementarity requires, and so seems very
promising.  But one must of course
be suspicious because light-cone gauge fields are not really local: even
in field theory there are nonlocal commutators.  The only gauge-invariant
observable which is available in string theory is the S-matrix itself,
but this should be enough: one can prepare an off-shell field by colliding
two on-shell packets.  Thus the commutator, if it is not a gauge artifact,
should imply some sort of action-at-a-distance in the S-matrix as
well.\footnote{This is the point at which the various authors of ref.~4
begin to differ in their interpretation of the result.}
Thus far no indication of this has been found.  Consider for example the
four-point amplitude, with particles 2 and 3 in the infalling frame and 1
and 4 on the stretched horizon.  According to the above discussion
\begin{equation}
p_{1,4}^+ = O(e^{-\omega}), \qquad
p_{1,4}^- = O(e^{\omega}). \label{boost}
\end{equation}
This four-point amplitude is rather
nonlocal off-shell, like the commutator, but on-shell the Virasoro-Shapiro
amplitude becomes
\begin{equation}
\delta({\mbox{$\sum$}} \vec p_{i}) \delta(p_1^- + p_4^-) \delta(p_2^+
+ p_3^+)\,  (\vec p_1 + \vec p_4)^{-2}
(p_1^- p_2^+)^{2 - (\vec p_1 + \vec p_4)^2 \alpha'/4}.
\end{equation}
This is independent of $p_{1,4}^+$ and so a delta-function shock wave
in $x_{1,4}^-$, just as in field theory.  The one difference from field
theory is the appearance of $(\vec p_1 + \vec p_4)^2$ in the exponent.
This produces the transverse spreading, and also an interesting
nonlocality in the
$x^-$ direction, but not the necessary nonlocality in the $x^+$ direction.
The same appears to be true of more complicated amplitudes.\footnote
{But see also ref.~14.  The processes considered in this paper are not
`nice' in the sense about to be defined, so I do not know if they are
relevant.  They are also highly suppressed.}

To conclude this section, let us mention one point arising in the
analysis of `nice' processes, where some momenta (`infalling')
are held fixed while the remainder are scaled as in
eq.~(\ref{boost}).
This is the Regge region, which has been
analyzed in some detail in ref.~15.  We will give a simplified analysis
applicable at tree level. The amplitude is dominated by the region where
the infalling vertex operators come together.  The case of two tachyon
vertex operators illustrates the main point.  The relevant OPE is
\begin{eqnarray}
&&\int d^2z :\!e^{ip_2 \cdot X(z)}\!: \, :\!e^{ip_3 \cdot X(0)}\!:
\ \sim
\ \int d^2z :\!e^{i (p_2 + p_3) \cdot X(0) + i p_2 \cdot
(z\partial + \bar z \bar\partial) X(0)} \!: |z|^{p_2
\cdot p_3 \alpha'} \\
&&\qquad =  2\pi \frac{\Gamma(p_2 \cdot p_3 \alpha' / 2)}
{\Gamma(1 - p_2 \cdot p_3 \alpha' / 2)}
:\! \Bigl[p_2 \cdot \partial X(0)  p_2 \cdot \bar\partial X(0)
\Bigr]^{-1-p_2
\cdot p_3 \alpha' / 2} e^{i (p_2 + p_3) \cdot X(0)} \!:\, .\nonumber
\end{eqnarray}
This is a curious result, involving fractional powers in the vertex
operator.  The general nice process factorizes at
tree level using the OPE in this way.

\section{Conclusion}

A summary, with additional remarks:
\begin{enumerate}
\item
Following low energy field theory until it breaks down leads to a state
with Hawking radiation in a highly mixed state, correlated with a
near-Planckian black hole with a large number of internal states.
\item
Nevertheless it may be that low energy ceases to be valid and stringy
effects become important sooner, when a nonlocal rather than local
invariant becomes large.
\item
A straightforward evaluation of the light-cone commutator shows just this
effect.  However, it remains to be seen whether this is a gauge artifact.
\item
Black hole complementarity is a logical possibility , and survives simple
attempts to prove it inconsistent.$^{16}$  In particular, one might worry
that if the external fields in fig.~2b are truly in a pure state, then the
state in fig.~2a cannot be the vacuum but must have real high-energy
quanta, a possibility which is generally regarded as unacceptable.  I do
not believe that this point is settled, but it appears to me that the
correlations needed are between fields highly spread out in time (the
black hole lifetime) and space, and will not lead to large local effects.
\end{enumerate}
To conclude, it is an exciting possibility that stringy behavior might
appear in a regime where it was not expected, and deserves further
attention.

\section{Acknowledgements}
I would like to thank S. Chaudhuri, S. Giddings, D. Lowe, A. Strominger, L.
Susskind, L. Thorlacius, J. Uglum, E. Verlinde, and H. Verlinde for
discussions.  This work was supported by NSF grants PHY-91-16964 and
PHY-94-07194.


\section{References}
\begin{thebibliography}{9}
\bibitem{Hawk} S.~Hawking, {\it Phys. Rev.} {\bf D14} (1976) 2460.
\bibitem{tHooft} G.~'t~Hooft, {\it Nucl. Phys.} {\bf B256} (1985) 727;
{\it Nucl. Phys.}
{\bf B335} (1990) 138; {\it Phys. Scr.} {\bf T36} (1991) 247.
\bibitem{STU} L.~Susskind, L.~Thorlacius, and J.~Uglum, {\it Phys.
Rev.} {\bf D48} (1993) 3743.
\bibitem{LPSTU} D.~Lowe, J.~Polchinski, L.~Susskind, L.~Thorlacius, and
J.~Uglum, {\it Black Hole Complementary vs.~Locality,} preprint
NSF-ITP-95-47, hep-th/9506138.
\bibitem{wald} R.~M.~Wald, {\it Space, Time and Gravity: the Theory of the
Big Bang and Black Holes} (Chicago U., EFI, 1992).
\bibitem{page} D.~N.~Page, {\it Phys. Rev. Lett.} {\bf 71} (1993) 1291.
\bibitem{polstrom} J.~Polchinski and A.~Strominger, {\it Phys.
Rev.} {\bf D50} (1994) 7403.
\bibitem{strom}  A.~Strominger, {\it Unitary Rules for Black Hole
Evaporation,} preprint UCSBTH-94-34, hep-th/9410187.
\bibitem{transv} L.~ Susskind, {\it Phys. Rev. Lett.} {\bf 71} (1993)
2367; {\it Phys. Rev.} {\bf D49} (1994) 6606.
\bibitem{diff}  A.~Mezhlumian, A.~Peet, and L.~Thorlacius,
{\it Phys. Rev.} {\bf D50} (1994) 2725.
\bibitem{super} L.~Susskind, {\it The World as a Hologram}, preprint
SU-ITP-94-33, hepth/9409089.
\bibitem{lccom} E.~Martinec, {\it Classical and Quantum Gravity} {\bf 10}
(1993) L187; D.~A.~Lowe, {\it Phys. Lett.} {\bf B326} (1994) 223.
\bibitem{LSU} D.~A.~Lowe, L.~Susskind, and J.~Uglum, {\it Phys. Lett.} {\bf
B327} (1994) 226.
\bibitem{lowe} D.~A.~Lowe, {\it The Planckian Conspiracy: String Theory
and the Black Hole Information Paradox,} preprint UCSBTH-95-11,
hep-th/9505074.
\bibitem{ACV}  D. Amati, M. Ciafaloni, and G. Veneziano,
Phys. Lett. {\bf B216}, 41 (1989); {\it Nucl. Phys.}
{\bf B403} (1993) 707.
\bibitem{ST} L.~Susskind and L.~Thorlacius, {\it Phys. Rev.}
{\bf D49} (1994) 966.

\end{thebibliography}
\end{document}